\documentstyle[prl,preprint,aps,epsf]{revtex}
\begin{document}
%\preprint{KNTP-99-02}
\bibliographystyle{plain}
\title{Supersphere}
\author{D. K. Park$^{1}$\footnote{e-mail:
dkpark@hep.kyungnam.ac.kr},
S. Tamaryan$^{2}$\footnote{e-mail: sayat@moon.yerphi.am} and
H. J. W. M\"{u}ller-Kirsten$^{3}$\footnote{e-mail:
mueller1@physik.uni-kl.de}
}
\address{
1.Department of Physics, Kyungnam University, Masan, 631-701, Korea\\
2.Theory Department, Yerevan Physics Institute, Yerevan-36, 375036, Armenia\\
3.Department of  Physics, University of Kaiserslautern, 
  67653 Kaiserslautern, Germany 
}
\date{\today}
\maketitle

\begin{abstract}
The spherical D2-brane solution is obtained without RR external
background. The solution is shown to preserve $(1/4)$ supersymmetries.
The configurations obtained depend on the integration constant $R_0$.
For $R_0 \neq 0$ the shape of the solution is a deformed sphere.
When, however, $R_0 = 0$, the D2-brane system seems to exhibit a 
brane-anti-brane configuration.
\end{abstract}
\maketitle
% \tableofcontents
% \listoffigures
\newpage
% %
%\section{Introduction}
After publication of Ref.\cite{call97}  on the description
of the D3-brane physics by Born-Infeld (BI) theory,
 much attention was  paid to 
the stability problem of the D-brane. In Ref. \cite{empa97}
 the 
stable cylindrical D2-brane was considered in which the
 electric worldvolume field and the 
non-trivial RR background are involved. In this case the 
potential allows
 only one locally stable vacuum,
and thus the fundamental string becomes a
 cylindrical D2-brane via  
quantum tunneling. The tunneling process 
has been examined in more detail  and 
generalized in Ref. \cite{park00-1}. In particular,
 it was found in  Ref. \cite{park00-1} 
that the RR background plays an important role in deciding
 the type of the quantum-classical phase transition. 

If one considers a pure magnetic worldvolume field 
instead of the  electric field,
the RR background allows  the D0-particles
 to expand to the spherical
D2-brane  in analogy to the dielectric effect
of Ref.  \cite{myers99}. This dielectric
effect has been  extended to the tubular case \cite{hyaku01}
 and has  also been
re-examined from the viewpoint of  semiclassical quantum 
tunneling \cite{park01-1}. 

More recently, a  new mechanism to obtain
 a stable tubular D2-brane without 
RR background was observed in Ref.\cite{mat01}
for the case when both
electric and magnetic worldvolume fields are turned on.
 In this case  
 angular momentum is responsible for the stability of the tubular
D2-brane,  and the Bogomol'nyi-Prasad-Sommerfield
(BPS) nature of the solution
indicates that the tubular D2-brane is generated by a
 `blowing-up' of the 
D0-particle-charged fundamental string. Since the final tubular D2-brane 
preserves  $1/4$ supersymmetries, it has been  named `supertube'.

Subsequently, the supertube solution was
 found in supergravity theory, and 
it was  found that the existence of the upper
 bound of the angular momentum
is related to the global violation of causality \cite{empa01}.
 The supertube 
solution has also  been examined  from the viewpoint
 of  matrix 
theory \cite{bak01},
 and has been  extended to the tubular D2-brane with 
arbitrary cross section in Ref. \cite{mat02}
 and to the  brane-anti-brane system in Ref. \cite{cho02}.
Furthermore, the supersymmetric D3-brane and their dualities
 have been  examined
 recently in Ref. \cite{tama02}. 
Although much attention was  paid to the 
supersymmetric tubular case, upon our knowledge  no further 
investigation has been devoted to  
 a supersymmetric sphere which has the  $S^2$ topology.

Here  we   examine the spherical supersymmetric 
D2-brane, or  so-called `supersphere'  by exploring the BI action
\begin{equation}
\label{biaction1}
S_{BI} = - \int d^3 \sigma \sqrt{-det(g_{ab} + F_{ab})}
\end{equation}
where $g_{ab}$ is the induced metric on the brane and $F_{ab}$ is the
worldvolume field strength. Moreover we set $T_2=1$ and
$2\pi\alpha^{\prime} = 1$.
As target space we choose a flat spacetime with   spherical coordinates
\begin{equation}
\label{bulk2}
ds^2 = -dT^2 + dR^2 + R^2 (d\Theta^2 + \sin^2 \Theta d\Phi^2)
      + ds^2(E^6),
\end{equation}
and define  worldvolume coordinates  as $t$, $\theta$, $\phi$,  where
\begin{equation}
\label{wvolume2}
t=T\, ,\qquad \theta = \Theta \, ,\qquad \phi=\Phi\, .
\end{equation}
Here we assume that $R$, which describes the
fluctuations of the brane,
 is   generally a  function of $t$, $\theta$, 
and $\phi$. Then the computation of the induced metric is straightforward.
Defining the worldvolume field strengths  as
\begin{equation}
\label{wfield2}
E_{\theta} = F_{t \theta}\, ,\qquad E_{\phi} = F_{t \phi}\, ,\qquad 
B = F_{\theta \phi}\,   ,
\end{equation}
the BI action (\ref{biaction1}) reduces to
\begin{equation}
\label{action2}
S_{BI} = \int dt d\theta d\phi {\cal L}_{BI}
\end{equation}
where
\begin{eqnarray}
\label{bilagrangian1}
{\cal L}_{BI}&=&-\bigtriangleup_s = 
-\Bigg[ R^2(R^2 \sin^2 \theta + R_{\theta}^2 \sin^2 \theta + R_{\phi}^2
       -E_{\theta}^2 \sin^2 \theta - E_{\phi}^2) + B^2
                                                \\    \nonumber 
& &\hspace{2.0cm}        
       - (E_{\phi} R_{\theta} - E_{\theta} R_{\phi})^2 + 
       2 B (E_{\phi} R_{\theta} - E_{\theta} R_{\phi}) \dot{R}
      -(R^4 \sin^2 \theta + B^2) \dot{R}^2 \Bigg]^{\frac{1}{2}}
\end{eqnarray}
In Eq.(\ref{bilagrangian1}) $\dot{R} = \partial_t R$ and 
$R_{\xi} \equiv \partial_{\xi} R$.

We now  take  $R_{\phi} = E_{\phi} = 0$
 for simplicity. Then ${\cal L}_{BI}$
in Eq. (\ref{bilagrangian1})  simplifies to
\begin{equation}
\label{bilagrangian2}
{\cal L}_{BI} = -\bigtriangleup_s = - 
\sqrt{R^2 \sin^2 \theta (R^2 + R_{\theta}^2 - E_{\theta}^2) + B^2 - 
(R^4 \sin^2 \theta + B^2) \dot{R}^2}.
\end{equation}
Since we assumed the time-dependence of $R$, $R$ is not 
 merely  a parameter,
{\it i.e.} it is a dynamical variable. Varying the BI action
$S_{BI}$ with respect to $R$, one can determine
 the dynamics of $R$ governed by
\begin{equation}
\label{dynarr}
\frac{\partial}{\partial t}
\frac{\dot{R}(R^4 \sin^2 \theta + B^2)}{\bigtriangleup_s} =
\frac{\partial}{\partial \theta}
\frac{R^2 \sin^2 \theta R_{\theta}}{\bigtriangleup_s} - 
\frac{2R^3 \sin^2 \theta (1 - \dot{R}^2) + R \sin^2 \theta (R_{\theta}^2
      - E_{\theta}^2)}{\bigtriangleup_s}. 
\end{equation}
The remaing equations of motion derived by varying $S_{BI}$ with respect
to the worldvolume gauge fields are
\begin{eqnarray}
\label{eqsmotion1}
\frac{\partial}{\partial \theta} \Pi_{\theta,E}&=&
\frac{\partial}{\partial \theta} \Pi_{\theta,M} = 0,  \\  \nonumber
\frac{\partial}{\partial t} \Pi_{\theta,E}&+&
\frac{\partial}{\partial \phi} \Pi_{\theta,M} = 0,
\end{eqnarray}
where
\begin{eqnarray}
\label{defpi1}
\Pi_{\theta,E}&=&\frac{R^2 \sin^2 \theta E_{\theta}}{\bigtriangleup_s},
                                                        \\    \nonumber
\Pi_{\theta,M}&=&\frac{(1 - \dot{R}^2) B}{\bigtriangleup_s}.
\end{eqnarray}
In the  static limit  Eq.(\ref{dynarr}) reduces to
\begin{equation}
\label{stadynarr}
\frac{\partial}{\partial \theta}
\frac{R^2 \sin^2 \theta R_{\theta}}{\bigtriangleup_s} = 
\frac{R \sin^2 \theta (2R^2 + R_{\theta}^2 - E_{\theta}^2)}{\bigtriangleup_s},
\end{equation}
and Eq.(\ref{eqsmotion1}) becomes
\begin{equation}
\label{staeqsmotion}
\frac{\partial}{\partial \theta} \Pi_{\theta,E} = 
\frac{\partial}{\partial \theta} \Pi_{\theta,M} = 
\frac{\partial}{\partial \phi} \Pi_{\theta,M} = 0,
\end{equation}
where
\begin{eqnarray}
\label{defpi2}
\bigtriangleup_s&=&\sqrt{R^2 \sin^2 \theta (R^2 + R_{\theta}^2 - E_{\theta}^2)
                           + B^2},            \\   \nonumber
& &\Pi_{\theta,E} = \frac{R^2 \sin^2 \theta E_{\theta}}{\bigtriangleup_s},
                                             \\   \nonumber
& &\Pi_{\theta,M} = \frac{B}{\bigtriangleup_s}.
\end{eqnarray}
If one considers a static theory from 
the beginning, the radial equation
(\ref{dynarr}) is not an equation of motion because $R$ is not a dynamical
variable. Thus, the solutions of the static theory do not need to satisfy it.
In the following, however, we
 will show that our solutions preserving $1/4$
supersymmetries satisfy the radial equation automatically
 in  a special limit.

From the definition of $\Pi_{\theta,E}$ in Eq.(\ref{defpi2}), 
$E_{\theta}$ is expressed in terms of 
$B$ and $\Pi_{\theta,E}$ as follows;
\begin{equation}
\label{relation1}
E_{\theta} = \frac{\Pi_{\theta,E}}{R \sin \theta}
\sqrt{\frac{B^2 + R^2 \sin^2 \theta (R^2 + R_{\theta}^2)}
           {\Pi_{\theta,E}^2 + R^2 \sin^2 \theta} }.
\end{equation}

Before solving the equations of motion, let us consider the energy of this 
D2-brane system. Performing the Legendre transform and using the Gauss
constraint, {\it i.e.} $\partial_{\theta} \Pi_{\theta,E} = 0$, one
obtains for the energy
\begin{equation}
\label{energy1}
{\cal E}_{BI} = \int d\theta d\phi {\cal H}
\end{equation}
where
\begin{equation}
\label{hamil1}
{\cal H} = \Pi_{\theta,E} E_{\theta} - {\cal L}_{BI}
= \frac{1}{R \sin \theta}
\sqrt{(\Pi_{\theta,E}^2 + R^2 \sin^2 \theta) 
      [B^2 + R^2 \sin^2 \theta(R^2 + R_{\theta}^2)]}.
\end{equation}

Now, we have to find the configurations $R(\theta)$, $E_{\theta}(\theta)$, and
$B(\theta, \phi)$ which minimize the energy (\ref{energy1}).
 In order to find
these configurations, we have to rely on 
the equations of motion. In fact, however,
Eq.(\ref{staeqsmotion}) does not completely fix the minimal configurations. 
For example, the equations of motion for $\Pi_{\theta,M}$
in 
Eq.(\ref{staeqsmotion}) can be solved by
 choosing $E^2 = R^2 + R_{\theta}^2$
or by choosing differently 
$B = b R \sin \theta \sqrt{R^2 + R_{\theta}^2 - E_{\theta}^2}$ where $b$ is a
constant. Each choice yields different solutions. 
This means we need another 
constraint which makes it possible to fix the minimal configurations. We believe
this additional constraint is obtained from the supersymmetric argument.

The number of supersymmetries preserved by any D2-brane configuration is the 
number of independent Killing spinors $\epsilon$ satisfying
\begin{equation}
\label{susy1}
\Gamma \epsilon = \epsilon
\end{equation}
where $\Gamma$ is a matrix defining the $\kappa$-symmetry 
transformation on the
worldvolume of the  D2-brane and is given  in our case as
\begin{equation}
\label{susy2}
\Gamma = \bigtriangleup_s^{-1} (\gamma_{t\theta\phi} + E_{\theta} \gamma_{\phi}
\Gamma_{\natural} + B \gamma_t \Gamma_{\natural}).
\end{equation}
In Eq.(\ref{susy2}) $\Gamma_{\natural}$ is a
 constant matrix with unit square
which anticommutes with all ten spacetime Dirac matrices $\Gamma_X$ and
$(\gamma_t, \gamma_{\theta}, \gamma_{\phi})$ are
 the induced worldvolume Dirac
matrices computed from $\Gamma_X$ as follows;
\begin{eqnarray}
\label{susy3}
\gamma_t&=&\Gamma_T,        \\   \nonumber
\gamma_{\theta}&=&R \Gamma_{\Theta} + R_{\theta} \Gamma_R,   \\  \nonumber
\gamma_{\phi}&=&R \sin \theta \Gamma_{\Phi} + R_{\phi} \Gamma_R.
\end{eqnarray}
Defining the Killing spinor $\epsilon$ in Eq.(\ref{susy1}) as
\begin{equation}
\label{susy4}
\epsilon = M_+ \epsilon_0,
\hspace{2.0cm}
M_{\pm} \equiv \exp \left( \pm \frac{1}{2} \Phi \Gamma_{R\Phi} \right)
\end{equation}
where $\epsilon_0$ is a  $32$-component constant
 spinor, the Killing
equation (\ref{susy1}) is decomposed into
\begin{eqnarray}
\label{susy5}
& &M_{+} \left[R R_{\theta} \sin \theta \Gamma_{TR\Phi} + R_{\theta} R_{\phi}
               \Gamma_T + B \Gamma_T \Gamma_{\natural} - \bigtriangleup_s \right]
                                                         \epsilon_0 +
                                                                    \\  \nonumber
& & \hspace{4.0cm}
 M_- \gamma_{\phi} \Gamma_{\natural} (R \Gamma_T \Gamma_{\natural} + E_{\theta})
                                                         \epsilon_0 = 0,
\end{eqnarray}
which yields two criteria 
\begin{eqnarray}
\label{susy6}
& &(R \Gamma_T \Gamma_{\natural} + E_{\theta}) \epsilon_0 = 0,
                                                              \\  \nonumber
& &(R R_{\theta} \sin \theta \Gamma_{TR\Phi} + B \Gamma_T \Gamma_{\natural})
                                              \epsilon_0 = 
\bigtriangleup_s \epsilon_0
\end{eqnarray}
for the preservation of supersymmetry.

Comparing with  the supertube case of  Ref. \cite{mat01}, 
one can conjecture from the 
first of  Eqs. (\ref{susy6}) that $E_{\theta} = \pm R$, and $\epsilon_0$
satisfies 
\begin{equation}
\label{susy7}
\Gamma_T \Gamma_{\natural} \epsilon_0 = - sgn(E_{\theta}) \epsilon_0.
\end{equation}
Since $E^2 = E_{\theta}^2 / R^2 + E_{\phi}^2 / R^2 \sin^2 \theta$
 in spherical
coordinates, the  supersymmetry
 preserving condition $E_{\theta} = \pm R$ yields
$E^2 = 1$,  which is a usual BPS condition\cite{mat01}.

Using $E_{\theta} = \pm R$ the second of  Eqs. (\ref{susy6}) becomes 
\begin{equation}
\label{susy8}
(R R_{\theta} \sin \theta \Gamma_{TR\Phi} + B \Gamma_T \Gamma_{\natural})
                                              \epsilon_0 =
\sqrt{R^2 R_{\theta}^2 \sin^2 \theta + B^2} \epsilon_0.
\end{equation}
Letting 
\begin{equation}
\label{susy9}
B = b R R_{\theta} \sin \theta
\end{equation}
where $b$ is some constant parameter, Eq.(\ref{susy8}) reduces to
\begin{equation}
\label{susy10}
(\Gamma_{TR\Phi} + b \Gamma_T \Gamma_{\natural}) \epsilon_0 = 
\sqrt{1 + b^2} \epsilon_0.
\end{equation}
Since the two conditions (\ref{susy7})
 and (\ref{susy10}) are compatible with each
other, the configurations satisfying (\ref{susy9})
 and $E_{\theta} = \pm R$
imply   a spherical D2-brane which preserves $1/4$ supersymmetries.
 Thus, the
remaining problem we should clarify is to check whether the supersymmetric
criteria (\ref{susy9}) and $E_{\theta} = \pm R$ are consistent with 
the  equations 
of motion or not. 

To check the consistency we first note that the 
supersymmetric criteria reduce
$\bigtriangleup_s$ to  $\sqrt{1 + b^2} B / b$,
 which automatically solves
$\partial_{\theta} \Pi_{\theta,M} =
 \partial_{\phi}\Pi_{\theta,M} = 0$. The 
remaining equation $\partial_{\theta} \Pi_{\theta,E} = 0$ 
is consistent with the
supersymmetric criteria if and
only if 
\begin{equation}
\label{compat1}
R^2 \sin \theta = E_0 R_{\theta},
\end{equation}
where $E_0$ is another constant parameter. Eq.(\ref{compat1}) has a solution
\begin{equation}
\label{solu1}
R = \frac{E_0}{|R_0 + \cos \theta|}
\end{equation}
where $R_0$ is an integration contant.
 The absolute value in  the denominator of  
Eq.(\ref{solu1}) is introduced to prevent $R$ from being negative. Thus, it is 
not necessary if we choose $R_0 > 1$. 
One can  show that the  solution
(\ref{solu1}) with the supersymmetric criteria is also consistent with
the relation (\ref{relation1}).

Fig. 1 shows the spherical D2-brane obtained from Eq.(\ref{solu1}) when 
$R_0 = 1.5$. Since Fig. 1 is apparantly different from tube, our solution
represents the `supersphere'. It is worthwhile noting that the supersymmetric 
criteria transform 
the radial equation (\ref{stadynarr}) into
$R_{\theta} = R \tan \theta$. Thus the
 solution (\ref{solu1}) with 
$R_0 = 0$ also solves this radial equation as stressed before. 
Fig. 2 is a spherical plot of 
solution (\ref{solu1}) with $R_0 = 0$. Fig. 2 seems to describe the 
brane-anti-brane system in the spherical coordinate. 

This conjecture gets more support
 if one computes the energy density 
for our solution. It is straightforward to show that the
 solution (\ref{solu1})
with the supersymmetric criteria
implies that  the energy density  (\ref{hamil1})
has the following form
\begin{equation}
\label{hamil2}
{\cal H} = T_2 \left(R \Pi_{\theta,E} + \frac{\sqrt{1 + b^2}}{b}
(2\pi \alpha^{\prime}) 
                     B \right)
\end{equation}
where $R$ in front of $\Pi_{\theta,E}$ appears to represent the string
length in the spherical coordinate as shown in the following. 
In Eq.(\ref{hamil2}) we have introduced $T_2$ and $2\pi \alpha^{\prime}$
explicitly to show that the energy of the D2-brane is 
\begin{equation}
\label{add1}
H = \int d\theta d\phi {\cal H} = n T_f \int R d\theta + 
m T_0 \frac{\sqrt{1 + b^2}}{b}
\end{equation}
where $T_f$ and $T_0$ are F-string tension and D0-brane tension respectively,
and the charges $m$ and $n$ are introduced by usual quantization
conditions $\Pi_{\theta,E} = n g$ and $\int B d\theta d\phi = 2 \pi m$. 
Of course, $\int R d\theta$ in Eq.(\ref{add1}) is a length of F-string.
When we take $b \rightarrow \infty$ limit in Eq.(\ref{add1}), $H$ becomes 
simply the sum of string and D0-brane energies like the supertube case in
Ref.\cite{mat01}. This result can be also conjectured from Eq.(\ref{susy10}).
If we take a $b \rightarrow \infty$ limit in this equation, Eq.(\ref{susy10})
becomes $\Gamma_{T} \Gamma_{\natural} \epsilon_0 = \epsilon_0$, which is 
same with that of the supertube case.

Apart from the factor of $R$ the expression  
${\cal H}$ in (\ref{hamil2}) is exactly 
the same as the  energy density for the 
super-D2-brane-anti-D2-brane tube in Ref.\cite{cho02}.
 This supports  our interpretation
of Fig. 2 as a brane-anti-brane pair.

%\begin{figure}
%\includegraphics{Fig1}% Here is how to import EPS art
%\caption{\label{fig:epsart} A figure caption. The figure captions are
%automatically numbered.}
%\end{figure}
%
%\begin{figure*}
%\includegraphics{Fig2}% Here is how to import EPS art
%\caption{\label{fig:wide}Use the figure* environment to get a wide
%figure that spans the page in \texttt{twocolumn} formatting.}
%\end{figure*}

%\begin{figure}
%\includegraphics{Fig1.ps}
%\caption{test1}
%\end{figure}

%\begin{figure}
%\includegraphics{Fig2.ps}
%\caption{test1}
%\end{figure}

Finally one can compute the worldvolume field strength using our solution
which reduces to 
\begin{equation}
\label{wfield3}
F = \frac{E_0}{R \sin \theta} dt \wedge dR + b R \sin \theta dR \wedge d\phi.
\end{equation}
Since $R \sin \theta$ is a radial distance in the cylindrical coordinate, 
Eq.(\ref{wfield3}) expresses the Coulomb-like electric field and the 
uniform magnetic field on the brane.

In the above  we have examined the spherical D2-brane which preserves 
$1/4$ supersymmetries. The shape of the D2-brane we  obtained 
 depends  on the integration constant $R_0$. For $R_0 \neq 0$
the shape of the D2-brane is a deformed sphere which we name `supersphere'.
However, when  $R_0 = 0$, the final D2-brane has a completely different 
topology, which we interpret as a brane-anti-brane system. It seems to be
interesting to examine the corresponding solutions   
in the context of supergravity
theory and the matrix theory.  

{\bf Acknowledgement:} This work was supported by the Kyungnam University
Research Fund, 2002.

\newpage

\centerline{\bf Figure Captions}

\vspace{0.9cm}

\noindent
{\bf Figure 1}

The spherical D2-brane which preserves $1/4$ supersymmetries. This figure
is obtained as a spherical plot with  $R_0 = 1.5$ using Eq.(\ref{solu1}).

\vspace{0.5cm}
\noindent
{\bf Figure 2}

The brane-anti-brane system derived from the limit of the  supersphere.
This figure is obtained by a spherical plot with
$R_0 = 0$ using Eq.(\ref{solu1}).

\newpage
\epsfysize=20cm \epsfbox{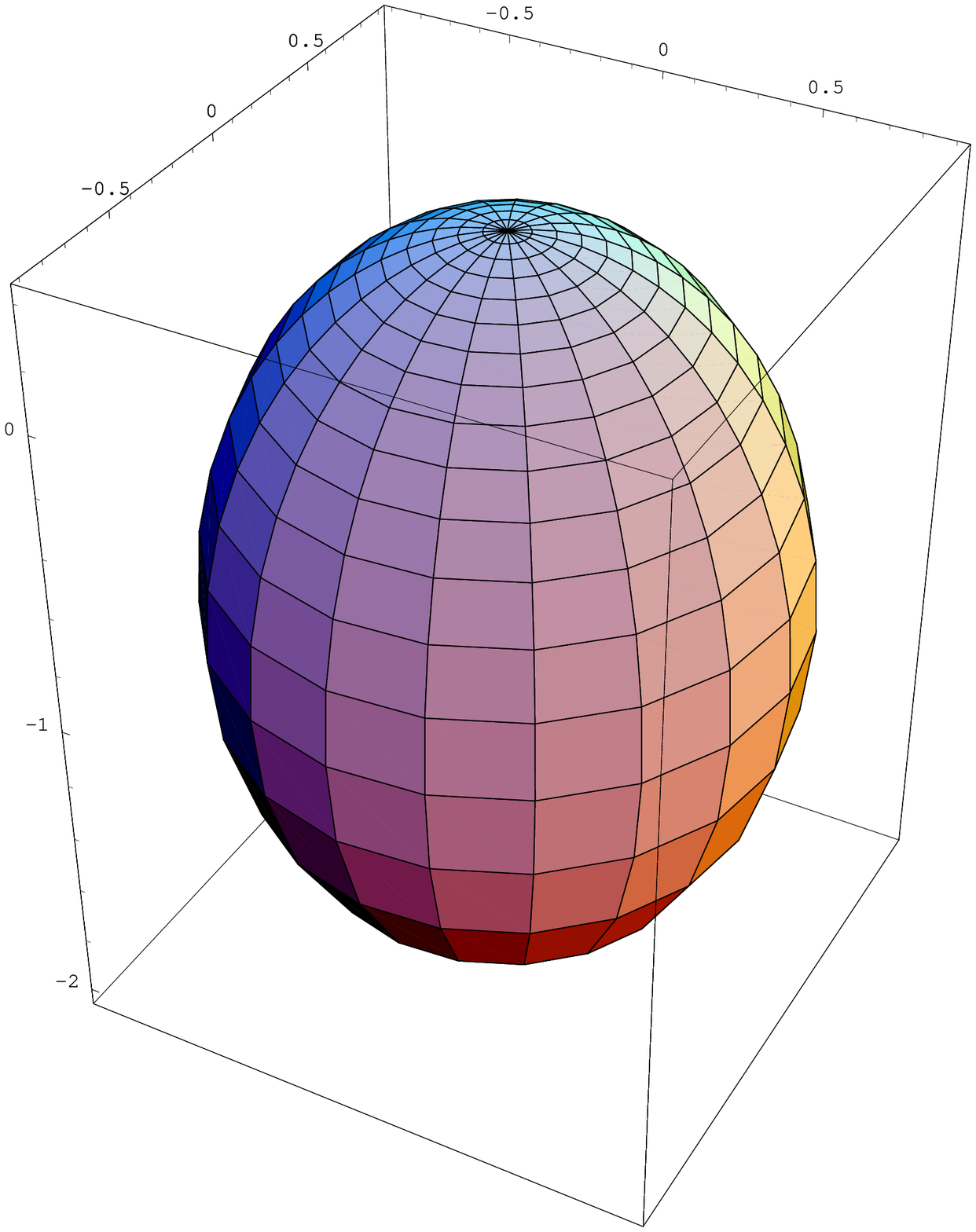}
\newpage
\epsfysize=20cm \epsfbox{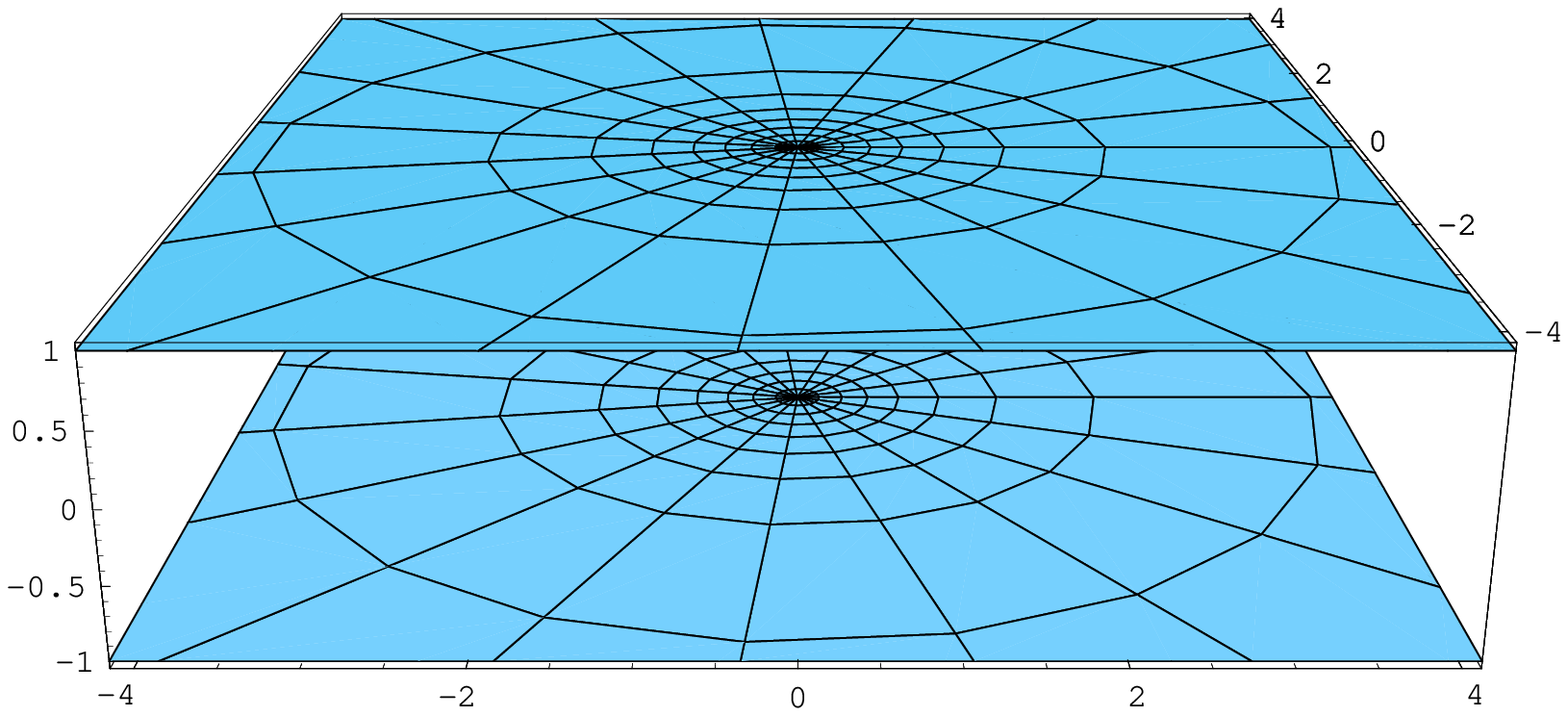}


\begin{thebibliography}{99}

\bibitem{call97} C. G. Callan and J. M. Maldacena, {\it Brane Dynamics From
the Born-Infeld Action}, Nucl. Phys. {\bf B 513} 198 [hep-th/9708147].

\bibitem{empa97} R. Emparan, {\it Born-Infeld Strings Tunneling to 
D-branes}, Phys. Lett. {\bf B 423} (1998) 71 [hep-th/9711106].

\bibitem{park00-1} D. K. Park, S. Tamaryan, Y.-G. Miao and 
H. J. W. M\"{u}ller--Kirsten, Tunneling of Born-Infeld Strings to 
D2-Branes, Nucl. Phys. {\bf B 606} (2001) 84 [hep-th/0011116].

\bibitem{myers99} R. C. Myers, {\it Dielectric Branes}, JHEP {\bf 9912}
(1999) 022 [hep-th/9910053].

\bibitem{hyaku01} Y. Hyakutake, {\it Torus-like Dielectric D2-Branes},
JHEP {\bf 0105} (2001) 013 [hep-th/0103146].

\bibitem{park01-1} D. K. Park, S. Tamaryan, and H. J. W. M\"{u}ller--Kirsten,
{\it D2-branes with magnetic flux in the presence of RR fields}, 
Nucl. Phys. {\bf B 635} (2002) 192 [hep-th/0111026].

\bibitem{mat01} D. Mateos and P. K. Townsend, {\it Supertubes}, Phys. Rev. Lett.
{\bf 87} (2001) 011602 [hep-th/0103030].

\bibitem{empa01} R. Emparan, D. Mateos, and P. K. Townsend, {\it Supergravity
Supertubes}, JHEP {\bf 0107} (2001) 011 [hep-th/0106012].

\bibitem{bak01} D. Bak and K. Lee, {\it Noncommutative Supersymmetric 
Tubes}, Phys. Lett. {\bf B 509} (2001) 168 [hep-th/0103148].

\bibitem{mat02} D. Mateos, S. Ng, and P. K. Townsend, {\it Tachyons, 
Supertubes and Brane/Anti-Brane Systems}, JHEP {\bf 0203} (2002)
016 [hep-th/0112054].

\bibitem{cho02} J. H. Cho and P. Oh, {\it Elliptic supertube and a 
Bogomol'nyi-Prasad-Sommerfield D2-brane-anti-D2-brane pair}, Phys. Rev. {\bf D 65}
(2002) 121901 [hep-th/0112106].

\bibitem{tama02} S. Tamaryan, D. K. Park, and H. J. W. M\"{u}ller--Kirsten,
{\it Tubular D3-branes and their Dualities} [hep-th/0209239].

\end{thebibliography}
\end{document}